# High Power on-chip Integrated Laser

Yair Antman[1,2,*], Andres Gil-Molina[1], Ohad Westreich[1], Xingchen Ji[1], Alexander L. Gaeta[2] and Michal Lipson[1]

[1]Department of Electrical Engineering, Columbia University, New York, New York 10027, USA
[2]Department of Applied Physics and Applied Mathematics, Columbia University, New York, New York 10027, USA
ml3745@columbia.edu
These authors contributed equally: Yair Antman and Andres Gil Molina

**The lack of high power integrated lasers have been limiting silicon photonics. Despite much progress made in chip-scale laser integration, power remains below the level required for key applications. The main inhibiting factor for high power is the low energy efficiency at high pumping currents, dictated by the small size of the active device. Here we break this power limitation by demonstrating a platform that relies on the coupling of a broad-area multimode gain to a silicon-nitride feedback chip, which acts as an external cavity. The feedback provided by the silicon-nitride chip is routed through a ring resonator and a single-mode filter, thus causing the otherwise multimode-gain chip to concentrate its power into a single highly-coherent mode. Our device produces more than 150 mW of power and 400 kHz linewidth. We achieve these high-performance metrics while maintaining a small footprint of 3 mm$^2$.**

Semiconductor lasers provide limited output power mainly due to the strong decrease in efficiency with increasing current in the gain medium. The power efficiency decreases due to the high current in the gain medium[1]. This limitation is especially strong at small form factors, such as the ones required to maintain a single transverse-mode output. In these high confinement gain waveguides both electrical and thermal resistance are high, causing the power dissipation to increase non-linearly with the laser's output power. The efficiency of lasers becomes prohibitively low for applications such as Light Detection and Ranging (LIDAR)[2] and WDM based on combs[3], that require hundreds of mW of power. Recently, hybrid integrated lasers were demonstrated where the gain element is coupled to a photonic integrated chip (PIC)[4–10], providing feedback to a single longitudinal mode. This design may enable both relatively high power and efficiency, however it requires a large footprint to lower the current density in the device[11].

Here we show single mode high power laser using a gain chip with less than 2 mm length. The gain medium is an off-the-shelf multimode Fabry-Perot Laser (MM-FPL) diode consisting of a 95 um wide active waveguide that supports hundreds of transverse and longitudinal modes. We ensure gain collapse to one mode by selectively providing higher loss to all other modes: loss to the longitudinal modes by using a silicon nitride (SiN) high-quality factor (high-Q) ring[12] and to the transverse modes using a tapered waveguide. We design the taper so that the coupling loss of all other modes is higher than that of the fundamental mode. In Fig. 1 one can see the gain medium with a width that is more than 20 times larger than the wavelength in the material, attached to a photonic chip with a tapered input waveguide and a high-Q ring. We design the ring to exhibit $Q_{loaded} = 0.5 \times 10^6$ with both input and drop ports. We terminate the bus waveguide of the drop port with a Sagnac loop mirror, designed for 100% reflection. The reflections from the Sagnac mirror propagate back into the MM-FPL through the high-Q ring and the single-mode waveguide, introducing feedback to only one longitudinal and one transverse mode, ensuring a highly coherent emission[13].

We achieve a single-mode output with more than 150 mW at a single frequency. In order to reach a self-injection locked state we align a single ring resonance to a longitudinal mode of the MM-FPL cavity and tune the phase of the input waveguide to ensure that the optical path along the waveguide induces constructive interference between reflection from the cavity and light exiting the FPL[14]. We achieve this through thermal tuning, by applying electrical power to the integrated platinum microheaters deposited on the cladding of the SiN chip. In Fig. 2a we show the measured output spectra in both the unlocked (top panel) and locked state (bottom panel). One can see that when the ring resonance frequency and waveguide phase are tuned so that the FPL is locked to a single ring resonance, the output power of the MM-FPL, otherwise distributed across a broadband multimode emission, collapses into a single peak in both spatial and spectral domains. In Fig. 2b we show the measured output power at the top panel and the simulated ones at the bottom panel (see Supplementary section 1), as a function of the pump current. One can see that the experimental results follow the expected trend from the simulations. The measured output power is high when compared to other single-mode laser diodes[15–17]. At pump current of 3.2 A we measure 150 mW single mode output.

We estimate the wall plug efficiency (Fig. 2c, top), given by $P_{out}/V_{pump}I_{pump}$ (where $P_{out}$, $V_{pump}$ and $I_{pump}$ are the measured output power, applied voltage and measured current), to be 4% at 150 mW output power. This wall plug efficiency corresponds to a modal collapse efficiency of 40%. These efficiencies are limited mainly by the coupling efficiency and sub-optimal transverse mode selectivity in the taper (see Supplementary section 5). The expected total wall plug efficiency and modal collapse, based on the measured threshold current and slope efficiency of the free-running laser, are up to 9% and 90% respectively.

We demonstrate up to 15 nm of continuous wavelength tunablity, as well as frequency pulling of up to 500 MHz. We achieve this tunability by tuning the spectral overlap between the ring modes and the MM-FPL modes. In Fig. 3a, we show the laser emission tuned within a band of up to 15 nm, by changing the ring resonance that overlaps with one of the MM-FPL modes. The wavelength may be fine-tuned by slightly varying the pump current in tandem with the ring resonance[18], thus enabling continuous tuning. We further demonstrate fine tuning of the emission wavelength by up to 500 MHz, by introducing a (small) detuning between the ring resonance and the MM-FPL mode, an effect known as frequency pulling[14]. We measured this fine tuning by beating the output of the laser with a fixed external laser, and detecting the interference with a high bandwidth detector (EOT ET-3010). The results, shown in Fig. 3b, were obtained by recording the output of the detector with an electrical spectrum analyzer.

We measured the linewidth of the locked laser to be 400 kHz. We use a delayed self-heterodyne setup, illustrated in Fig. 4a, where the output of the laser is split into two branches. In one branch it is delayed by a 500m single-mode fiber, and in the other it is shifted in frequency by an acousto-optic modulator[18]. We then detect the power of the recombined branches with a high bandwidth photodiode and record the signal using a real-time oscilloscope (Keysight MSOX4154A). In Fig. 4b, we show the Fourier transform of the recorded data. One can see that the 400 kHz linewidth exhibits a Lorentzian lineshape, indicating that the linewidth is dominated by the fundamental Schawlow-Townes limit. In addition, we plot the extracted single-sideband frequency noise (Fig. 4a) of the interference signal. We find that the frequency noise spectrum is flat across the entire measured bandwidth, further indicating that the linewidth is determined by the intrinsic properties of the system, dominated by spontaneous emission[19] and manifesting in a 400 kHz wide Lorentzian linewidth. See Supplementary section 4 for details on the analysis and further discussion on the lineshape of the laser.

Our results demonstrate a platform that breaks the trade-off between power and efficiency in a highly-scalable integrated laser. The power levels demonstrated in this study, mark an important step in unlocking ultra-high-power on-chip laser sources. Coherent emission of 150 mW is appropriate for important use cases such as comb generation for WDM. Our numerical calculations indicate (see Supplementary section 2) that the laser performance can be further increased by optimizing the taper design in the SiN chip, including an additional integrated filter and improving the packaging. We predict that more than 600 mW could be readily obtainable with this platform using the same gain chip. Since all the various components of the external cavity (ring, Sagnac reflector etc.) are implemented on the SiN PIC, there are no necessary modifications to the gain chip. Having identified the surface area of the gain as the source of the power-efficiency trade-off, it is important to note that we were able to significantly push the performance without sacrificing footprint. By extending the gain along the transverse axis the surface area was increased by a factor of 30, while maintaining the same total footprint as most standard single mode gain elements. The total footprint of our device, including both gain and SiN chip, is less than 3 mm$^2$. The platform allows us to leverage the high power handling of the large area gain while maintaining single mode operation output in a highly confined waveguide.

**Contributions**
Y.A, A.G.M, M.L. and A.L.G conceived and proposed the multimode gain collapse and its use for high power lasers. A.G.M. and O.W. designed the SiN devices, which were fabricated by X.J. O.W. and A.G.M. conceived and assembled the experimental interface between the active and passive chips. Testing and characterization were performed by A.G.M, Y.A. and O.W, and data analysis by Y.A. and A.G.M. Y.A. designed and performed the numerical model. Y.A and M.L. prepared the manuscript. A.G.M and M.L. edited the manuscript. M.L. and A.L.G supervised the project.

**Acknowledgements**
This work was supported by ARPA -E, PINE: Photonic Integrated Network Energy Efficient Datacenters program (DE-AR0000843); PIPES, Embedded Photonics ultra-bandwidth dense optical interconnect (EmPho) program (HR0011-19-2-0014); and ARO, Novel Chip-Based Nonlinear Photonic Sources from the Visible to Mid-Infrared program (W911NF2110286).

**Competing interests**
M.L., Y.A., J.X., and A.L.G are named inventors on US patent US 2022/0006260 A1 regarding the technology reported in this article.

**Methods**

**Device fabrication.** The detailed fabrication procedure of the SiN PIC can be found in ref.[12]. In this work, we used a 730 nm thick SiN layer, deposited using low pressure chemical vapor deposition (LPCVD). The width of the bus and ring waveguides is 1500 nm. The free spectral range of the ring is 200 GHz.

**Coupling between III/V multimode gain and SiN device.** The commercial MM-FPL (purchased at RPMC Lasers) used in this work was mounted on a standard C-mount. We connected the C-mount to a thermoelectrically cooled mount (Newport LDM-4409), set to a constant temperature of 23°C. Both the MM-FPL and SiN chip were placed on linear stages for mechanical alignment. The roll angle of the SiN was fine-tuned by an additional goniometer stage. We resolved the angle and elevation difference between the chips by placing the imaging system at a non-normal angle with respect to the faces of the chips. We wirebonded the SiN chip, and connected it to a custom built printed circuit board (PCB). We applied DC current to the heaters via the wirebonds with a digital power supply (Keithley 2230-30-3). The output waveguide is an inverse taper designed to produce a symmetric circular mode. We collimated the output with an aspheric lens, and coupled it to a single mode fiber with a fixed-focus fiber collimator.

**Numerical simulation.** We studied the effect of the external cavity on the output power of our device using a numerical model. The multimode laser we use in this work has extremely high power and WPE, of more than 30% at Watt-level output. Since self-injection locking inevitably introduces additional loss mechanism and inefficiencies, the single mode output will not retain the same level of power performance. Using the model, we quantify these mechanisms through numerical simulations. We show that despite the added loss induced by the external cavity and suboptimal modal collapse, the single-mode output emission remains at very high power levels. The full details of the model, its implementation and the main results or shown in Supplementary sections 1, 2 and 3.

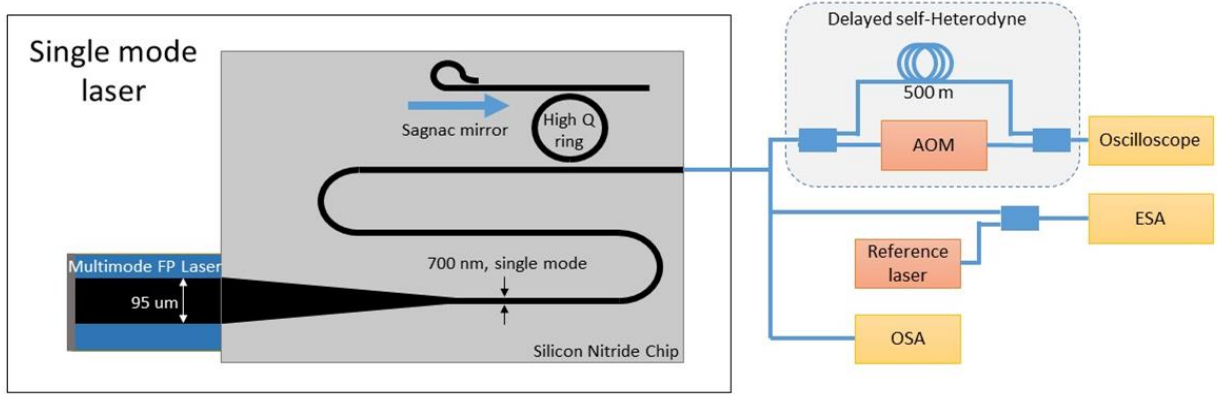

**Fig. 1 |** Schematic depiction of the experimental setup. A multimode Fabry-Perot Laser (MM-FPL) is coupled to a SiN chip, which provides feedback via a Sagnac loop mirror. The feedback is routed through a high quality-factor (high-Q) ring and an adiabatic taper. The spectrum of the laser is measured with an optical spectrum analyzer (OSA). Fine spectral tuning of the laser performed by frequency pulling is tested through interference with a reference laser, and measured with an electrical spectrum analyzer (ESA). The lineshape is extracted with a delayed self-heterodyne setup, where one copy of the laser is offset in frequency by an acousto-optic modulator (AOM) and the other is delayed by 500 m. The combined signal is detected and recorded with a real time oscilloscope.

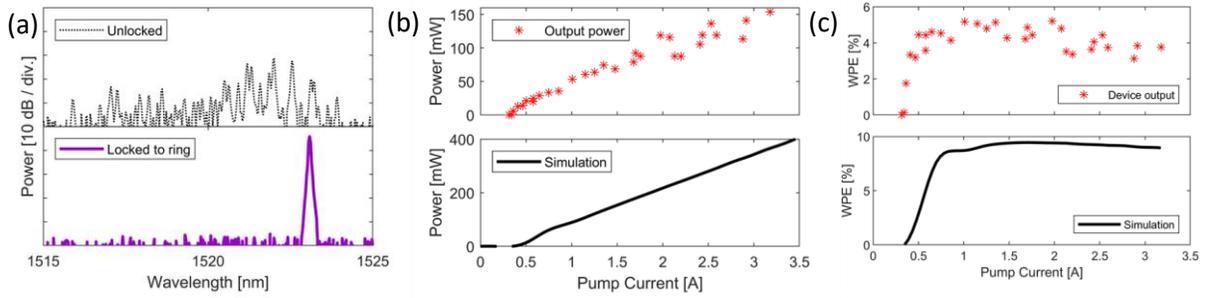

**Fig. 2 | Single mode emission and power characterization. a,** Single mode spectrum is obtained when coherent reflection from the Sagnac are routed to the MM-FPL through the ring (lower panel), compared with broad incoherent spectrum of the free-running laser (upper panel). **b,** Output power and **c,** wall-plug efficiency as a function of input current. Measurements taken at the output of the SiN device are indicated with red stars (upper panels) and corresponding simulation results plotted in solid black lines (lower panels).

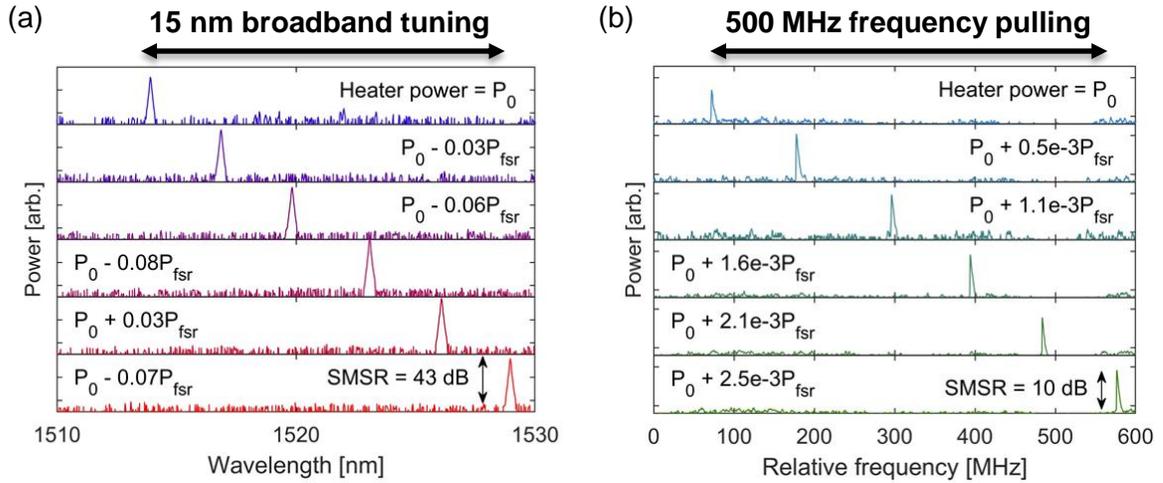

**Fig. 3 | Wavelength tuning.** Continuous broadband tuning: we select the lasing wavelength by aligning the frequency of a ring resonance to a specific MM-FPL mode and show up to 15 nm of tuning bandwidth. This is demonstrated by six OSA spectra shown in **a**. Frequency pulling: we shift the ring resonance while keeping the MM-FPL steady. By detecting the interference of our laser and a reference laser we demonstrate that the lasing frequency is pulled to that of the ring within a range of up to 500 MHz. The pulling range is shown in six ESA spectra **(b)**. Data in this figure were taken at pump current of 700 mA. SMSR – single-mode suppression ratio. $P_0 = 54$ mW is a nominal power and $P_{fsr} = 400$ mW is the heater power required to shift the ring resonance by a its free spectral range off 200 GHz.

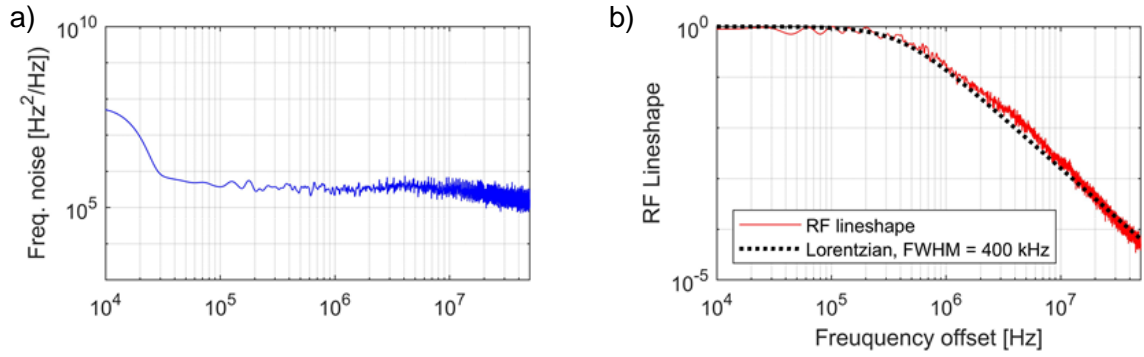

**Fig. 4 | Lineshape measurements.** We assess the lineshape of the laser with a delayed self-heterodyne setup. The power spectral density (PSD) of the frequency noise is calculated by extracting the phase of the sampled signal **(a)**. Direct observation of the RF lineshape **(b)** reveals a Lorentzian line with 400 kHz FWHM, in good agreement with the flat PSD of the frequency noise.

# Supplementary information of "High Power on-chip Integrated Laser"

Yair Antman[1,2,*], Andres Gil-Molina[1], Ohad Westreich[1], Xingchen Ji[1], Alexander L. Gaeta[2] and Michal Lipson[1]

[1]Department of Electrical Engineering, Columbia University, New York, New York 10027, USA
[2]Department of Applied Physics and Applied Mathematics, Columbia University, New York, New York 10027, USA

## 1. Simulation model

The simulation procedure is outlined in Fig S1. We analyze the interaction of the field with the gain point-by-point, on an $N \times M$ two-dimensional mesh. On each point in the mesh, we calculate the time-domain value of three quantities: left- and right- propagating E-fields, $E_r, E_l$ and the local population inversion $N$. The gain cavity is caped with two mirrors, a perfectly reflecting back mirror $r_b$ and a partially reflecting front mirror, $r_f$, taken to have 5% power reflectance. We simulate the interaction with an external cavity by inserting the fundamental mode power into a single mode waveguide $E_{bus}$, which is coupled to a ring resonator. Reflections from the ring are then injected back into the fundamental mode of the gain, and the system is propagated in time-domain until a steady-state is reached.

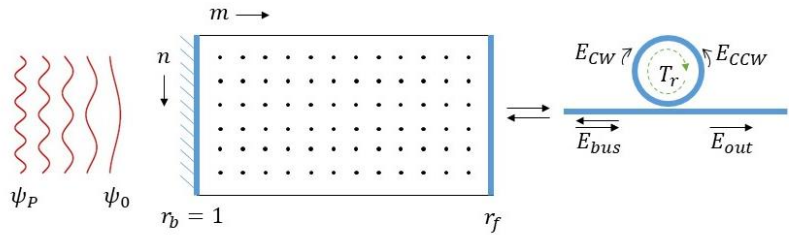

**Fig. S1** | Numerical simulation model. The gain medium is segmented, forming a two-dimensional $N \times M$ mesh, with the $n$ and $m$ indices indicating transverse and longitudinal coordinates, respectively. Transverse modes are denoted by $\psi_p$ and are depicted in red curves. The fundamental mode field is transmitted through the front mirror $r_f$ and is coupled into a ring resonator. Back-reflections from the ring are then transmitted back into the fundamental mode at the Mth column of the gain, forming an external cavity.

The local population inversion follows the rate equation[1]:

$$\frac{dN}{dt} = \frac{I}{e} - \frac{N}{\tau_s} - S \qquad (1)$$

In (1), $I$ is the current per mesh point, $e$ is the electron charge, $\tau_s$ is the upper-state life time of the gain medium and $S$ is the time and space dependent photon emission rate. Using $N$, we may calculate the gain in each point:

$$g = (N - N_g)\theta \qquad (2)$$

Here, $N_g$ is population inversion at transparency and $\theta$ is the differential gain coefficient. The fields then evolve in time according to the following equation:

$$\frac{dE}{dt} = (g - \eta_g)E + \sqrt{\beta h\nu N}W(t) \tag{3}$$

where $\eta_g$ is the linear loss per unit time in the gain medium, $\beta$ is the spontaneous emission rate, $h\nu$ is the photon energy and $W(t)$ is a random process defined as:

$$W(t) = W_r(t) + iW_i(t) \tag{4}$$

with $W_r(t)$ and $W_i(t)$ are jointly Gaussian with zero mean and unity variance. On each simulation time step $dt$, after evaluating Eq. (3), we propagate the modes along the longitudinal axis by separation to eigen-modes:

$$E_r(n, m, t) = \sum_p A_{p,r}(m - 1, t)\psi_p(n)e^{-jk_p dz} \tag{5}$$

$$A_{p,r}(m, t) = \langle \psi_p(n), E_r(n, m, t) \rangle \tag{6}$$

Here, $\psi_p(n)$ is the set of transverse eigen-modes of the active waveguide, with corresponding propagation constants $k_p$ and $dz$ is the finite length interval between longitudinal mesh points. Propagation of $E_l$ is done is a similar manner, along the negative $m$ axis. We calculated the modes separately with a finite-difference solver.

Concurrently, we analyze the effect of the ring as an external cavity. The field in the bus is coupled to the fundamental mode in the $m$th column in the gain:

$$E_{bus,r}(t) = \eta_c e^{i\phi_e}\left[\langle\psi_0(n), E_r(n, M, t - dt)\rangle t_f + \eta_c e^{i\phi_e} r_f E_{bus,l}(t)\right] \tag{7}$$

$$E_l(n, M, t) = E_r(n, M, t)r_f + \eta_c e^{i\phi_e} t_f E_{bus,l}(t)\psi_0(n) \tag{8}$$

where $\eta_c$ is the coupling loss, $\phi_e$ is the external cavity phase delay and $t_f = \sqrt{1 - r_f^2}$. In this model we assume moderate coupling loss of $\eta_c = 0.8$. We then calculate the circulating fields in the ring by solving the coupled equations[2] for $E_{CW}(t)$ and $E_{CCW}(t)$. All fields and the population inversion are solved simultaneously in time-domain, and the results are taken after all transients have subsided and the system is at steady-state.

## 2. Power analysis of self-injection locked multimode gain

We now employ our model in analyzing the effect on power output when forcing a multimode gain to operate in a single-mode regime. We perform a multi-dimensional sweep of four parameters in the model: the internal Q-factor of the ring, the bus-ring coupling rate, the external cavity phase delay and the ring detuning. We find that in all instances of the sweep, the power at the output of the device $P_{out}$, defined as the spectral peak of $|E_{out}|^2$, is predicted by the following heuristic equation:

$$P_{out} \approx (\eta_c \kappa_{gain} T_{ext} \eta_{SM}) \times P_{free} \tag{9}$$

In this equation, $\kappa_{gain}$ is the overlap between the fundamental mode and the gain profile, $\eta_{SM}$ is the fraction of single-mode power out of the total power in the gain medium (modal collapse efficiency), and $T_{ext}$ is the transmission of the ring, defined as $|E_{out}|^2/|E_{bus}|^2$. $P_{free}$ is the total multimode power of the free-running laser. In Fig. S2a we show the output of the simulations, plotted against the prediction of Eq. (9). The prediction is found to be accurate within a 10% relative error.

The heuristic model sheds light on the power penalties imposed by self-injection locking of a multimode gain. The power is reduced when compared to the free-running case due to two main reasons:

a. The external cavity introduces more loss mechanisms, both technical (coupling losses, $\eta_c$) and fundamental (ring transmission, $T_{ext}$).
b. The gain medium is multimode, hence non-ideal modal collapse will result in some higher-order mode power (quantified by $\eta_{SM}$), which results in reduced single-mode power. In addition, the overlap between the fundamental mode and the gain profile will be less than unity, degrading the slope efficiency.

There exists a tradeoff between two of the parameters: $T_{ext}$ and $\eta_{SM}$. In order to draw the power of the multimode gain to TE$_0$, we must provide sufficient feedback to that mode. This is shown in Fig. S2b where we plot the modal collapse efficiency as a function of ring power reflection. However, stronger reflections from the ring inherently reduces the transmission. The product, $T_{ext} \times \eta_{SM}$, is shown in Fig. S2c, where we note that the power peaks when the reflection of the ring is close to 10%.

We may now provide an estimation for the single-mode power at the output of the device. As shown if Fig. S2c, the product $T_{ext} \times \eta_{SM}$ peaks at 0.6 under ideal tuning conditions. The Q factor of the ring in this simulation is $10^6$, and it is over-coupled with extinction of 70%. $\kappa_{gain}$ for the laser in this study is found to be 0.84. We assume coupling losses of $\eta_C = 0.8$. Multiplying the above parameters, we get:

$$P_{out} \approx 0.4 \times P_{free} \tag{10}$$

We may now compare the expected injection-locked single-mode output to the free-running multimode output of the laser. For Eq. (10), we conclude that under the above assumptions, when the phase and spectral detuning between the cavities is ideal, we may expect the injection-locked output to be 40% of multimode free-running one. With 1500 mW multimode power at 3A pump current, we expect up to 600 mW at the output of the single-mode SiN waveguide.

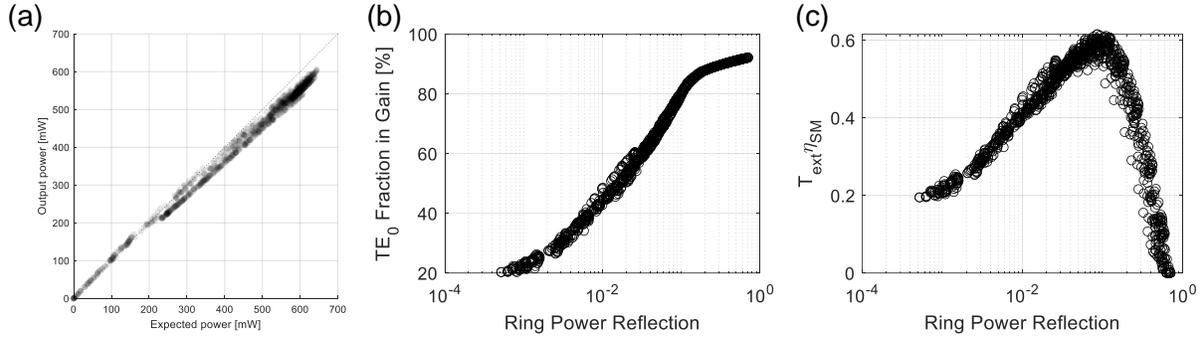

**Fig. S2 |** Simulation results, aggregated over a multi-dimensional parameter sweep. Each point represents a single realization, with a unique set of input parameters. **(a)** Single-mode power output, plotted against the heuristic prediction of Eq. (9). **(b)** Transverse modal collapse efficiency: power in the fundamental mode at the output of the gain medium, divided by the total multimode power ($\eta_{SM}$). It is plotted against the effective ring reflection: $|E_{bus,l}|^2/|E_{bus,r}|^2$. **(c)** The product $T_{ext} \times \eta_{SM}$, indicating the trade-off between high ring transmission and modal collapse efficiency.

## 3. Experimental power results and partial spectral collapse

We adapt our model to more closely match the experimental setup by setting the insertion loss to $\eta_c = 0.5$. Comparing the experimental results with the simulation, as shown in Fig. 2b, we see that they agree at low power, then diverge. The experimental peak power does not scale with $I - I_{th}$, as expected by the simulation. However, by measuring the total power in TE$_0$ rather than the spectral peak, (Fig. S3a) we find that it closely match the power found in simulations. As shown in Fig. S3c, the ratio of power contained in a single longitudinal mode, to the total power at the output of the SiN device, is decreasing at high pump currents. We learn from this that the transverse modal collapse stays highly effective at all pump currents, forcing the multimode gain to predominantly emit power at TE$_0$. In contrast, the dominance of a single longitudinal mode is not maintained at high power (Fig S3b), which explains the degradation in single-mode power at high current. This effect, namely lower spectral single-mode to side-lobe ratio at high power, was not predicted by our model, and is the subject of ongoing research.

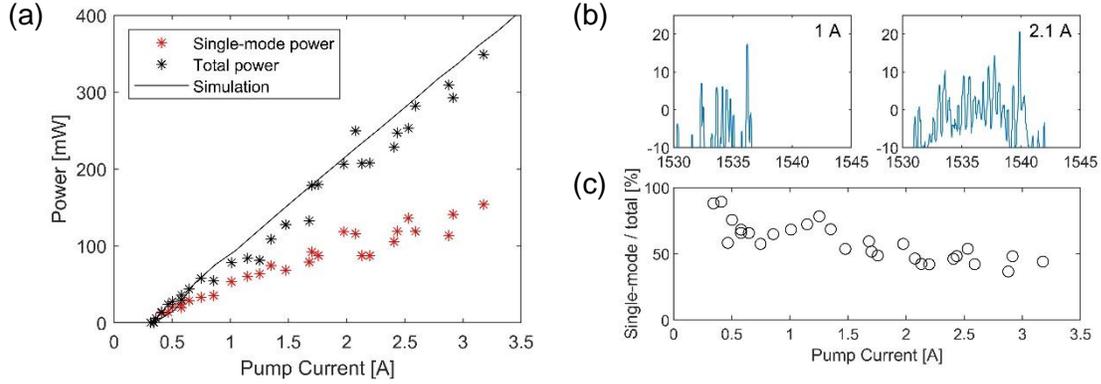

**Fig. S3 | (a)** Output power, taken at the output of the SiN device. Red stars indicate the coherent, single-mode power, measured at the spectral peak, with corresponding simulation result (black curve). Black stars indicate the total power at the output of the SiN waveguide. **(b)** Two sample spectra, taken with 1 A and 2.1 A pump current. Note that at high pump currents, the output consists of higher power off-peak spectral components. **(c)** Ratio between single-mode (red stars in **(a)**) and total power laser power (black stars in **(a)**). We observe that while the total power trends linearly with current, as expected by the simulations, the coherent power, in a single longitudinal mode, does not.

## 4. Frequency Noise Calculation

We measure the coherence of the laser using a delayed self-heterodyne setup, sampled in real time, as shown in Fig. 1b, with results in Fig. 4a. As we show in this section, the finite delay length of 500 m causes spurious fringes in the frequency noise spectrum, which we account for in order to extract the clean laser PSD. In delayed self-heterodyne, the detected voltage is:

$$v = C + \sin[\Omega t + \phi(t) - \phi(t - T_d)] \tag{11}$$

where $C$ is a DC component, $\Omega = 2\pi \times 100\ MHz$ is the offset frequency and $T_d$ is the delay time of the 500 m long fiber. Using a phase detection algorithm, we get the signal $\Gamma = \phi_1(t) - \phi_2(t)$, with $\phi_1(t) = \phi(t)$ and $\phi_2(t) = \phi(t - T_D)$ and corresponding PSDs $S_{\phi_1,\phi_1}$ and $S_{\phi_2,\phi_2}$. The PSD of $\Gamma$ is therefore:

$$S_{\Gamma,\Gamma} = S_{\phi_1,\phi_1} - 2Re\{S_{\phi_2,\phi_1}\} + S_{\phi_2,\phi_2} \tag{12}$$

Assuming $\phi$ is stationary, we may simply: $S_{\phi_1,\phi_1} = S_{\phi_2,\phi_2} = S_{\phi,\phi}$. The cross spectrum $S_{\phi_2,\phi_1}$ can be found using the cross correlation:

$$R_{\phi_2,\phi_1}(\tau) = <\phi(t)\phi(t - \tau - T_d)> = R_{\phi,\phi}(\tau - T_d) \tag{13}$$

$$S_{\phi_2,\phi_1} = FT\{R_{\phi,\phi}(\tau - T_d)\} = e^{j\omega T_d} S_{\phi,\phi} \tag{14}$$

And the calculated PSD is found to be:

$$S_{\Gamma,\Gamma}(\omega) = 2[1 - \cos(\omega T_d)]S_{\phi,\phi}(\omega) \tag{15}$$

From Eq. (15) we see that $S_{\Gamma,\Gamma}(\omega)$ is the PSD of the laser phase, $S_{\phi,\phi}(\omega)$, multiplied by a sinusoidal envelope. We calculate $S_{\Gamma,\Gamma}(\omega)$ using Welch's method[3] (see Fig 4, dashed black line), and divide the result by the appropriate sinusoidal envelope to reach $S_{\phi,\phi}(\omega)$ (Fig. 4, red solid line).

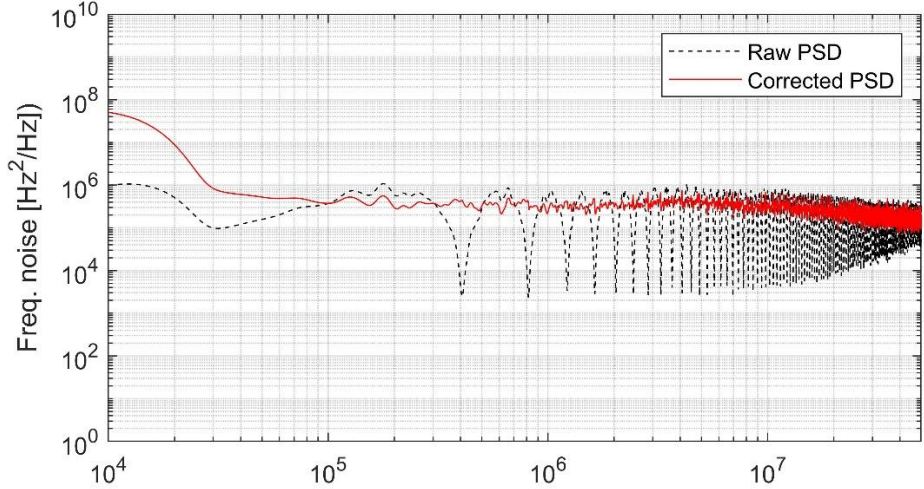

**Fig. S4 |** The PSD of the laser frequency noise (red) is extracted from the PSD of the detected delayed self-heterodyne signal (black dashed line) by removing the overlaying sinusoidal envelope.

## 5. Input linear taper

In our laser, a long horn-shaped taper is implemented to adiabatically convert from the 95 um wide waveguide to a narrow single-mode SiN waveguide. Under ideal conditions, the fundamental modes of the two waveguides are coupled with 100% efficiency, while no power transfer is allowed to higher order modes. We simulated our realization, a 2 mm long linear taper, using an EigenMode Expansion (EME) solver, and calculated the transfer matrix between the input and output sets of modes. In Fig. S5 we show the power transfer between the fundamental mode of the SiN waveguide and the 10 lowest order modes of the gain waveguide. We learn that about 80% of the power remains in the fundamental mode, which corresponds to 1 dB of added insertion loss. Additionally, a substantial amount of power is transferred to high order modes, and vice versa. Leakage between non-lasing high-order modes and the fundamental mode may result in a significant increase in spontaneous emission rates, which leads to broadening of the Lorentzian linewidth.

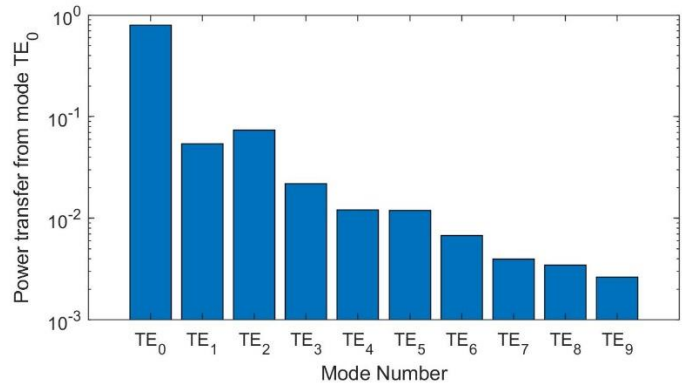

**Fig. S5 |** Power transfer into the multimode gain from a single-mode SiN waveguide.